\documentclass{aastex}
\usepackage{spr-astr-addons}

\begin{document}

\title{Extensive photometry of the intermediate polar V1033~Cas (IGR J00234+6141)}
\shorttitle{Photometry of IP V1033~Cas}
\shortauthors{Kozhevnikov}

\author{V. P. Kozhevnikov}
\affil{Astronomical Observatory, Ural Federal University, Lenin Av. 51, Ekaterinburg 
620083, Russia e-mail: valery.kozhevnikov@urfu.ru}

\begin{abstract} 

To measure the spin period of the white dwarf in V1033~Cas with high precision, we performed extensive photometry. Observations were obtained over 34 nights in 2017. The total duration of the observations was 143~h.We found that the spin period of the white dwarf is equal to $563.116\,33\pm0.000\,10$~s. Using this period, we derived the oscillation ephemeris with a long validity of 100 years. The spin oscillation semi-amplitude was stable and was equal to $95.5\pm1.3$~mmag. This is a very large semi-amplitude of the spin oscillation among intermediate polars, which have similar and lesser spin periods. This large semi-amplitude suggests that the system is noticeably inclined. The spin pulse profile was sinusoidal with high accuracy.  This may mean that the spin oscillation is produced by a single accretion curtain whereas the second accretion curtain may be obscured by the accretion disc. Despite the large amount of our observations, we did not detect sidebands. The semi-amplitudes of the undetected sideband oscillations do not exceed 10~mmag. The absence of sideband oscillations seems puzzling. We detected the orbital variability of V1033~Cas with a period of $4.0243\pm0.0028$~h and with a semi-amplitude of $55\pm4$~mmag. The orbital variability semi-amplitude seems large and also suggests that the system is noticeably inclined. Using our oscillation ephemeris and the times of spin pulse maximum obtained in the past, we found that the spin period is very stable. ${\rm d}P/{\rm d}t$ is most probably less than  $-4\times10^{-12}$.    This contradicts the assumption that the white dwarf in V1033~Cas is not spinning at equilibrium. Our spin period and our oscillation ephemeris can be used for further investigations of the stability of the spin period in V1033~Cas.  
\end{abstract}


\section{Introduction}
Cataclysmic variables (CVs) consist of a white dwarf that accretes material from a late type secondary star. Accretion can occur through a bright accretion disc when the white dwarf is non-magnetic, through a bright accretion column when the white dwarf is strongly magnetic and through a truncated accretion disc when the white dwarf is moderately magnetic. In the latter case, such CVs are called intermediate polars (IPs). Because of the many phenomena of variability, IPs are very interesting for detailed photometric investigations. In IPs, the magnetic white dwarf spins asynchronously with the orbital period of the system and therefore produces a rapid oscillation with the spin period. This oscillation is coherent and shows a stable period. Because this oscillation is produced not only in optical light but also in X-rays, X-ray reprocessing by the secondary star or the bright spot in the disc usually creates the oscillation with the beat period, $1/P_{\rm beat} = \omega-\Omega$, where $\omega=1/P_{\rm spin}$ and  $\Omega=1/P_{\rm orb}$. Often, such orbital sidebands are stronger than oscillations with the spin period. In addition, other orbital sidebands such as $\omega-2\Omega$ and $\omega+\Omega$ can be produced from amplitude modulation \citep{warner86}. So, multi-periodicity is a typical feature of IPs. Reviews of IPs are presented in \citet{patterson94, warner95, hellier01}.

The spin period modulation is the main characteristic of belonging to the IP class. Therefore, spin period measurements are a necessary condition for accepting a CV as an IP.  The spin period must be coherent and stable. In addition, the precise spin period and the precise oscillation ephemeris make it possible to perform an observational test of spin equilibrium either from direct measurements of the spin period or from pulse-arrival time variations by using a precise oscillation ephemeris. 

Although the IP nature of V1033~Cas was suspected 12 years ago, to date no precise spin period and precise oscillation ephemeris have been obtained. Indeed, for the first time, \citet{bikmaev06} noted that V1033~Cas shows an optical oscillation with a period of 570~s. But, they did not even define the possible error of this period. It seems unacceptable. Next, \citet{bonnet07} found that the period of the optical oscillation in V1033~Cas is equal to $563.53\pm0.62$~s. The precision of this period is low because this period was measured using data from a single observational night. Another observation of V1033~Cas was obtained in X-rays by \citet{anzolin09}. They found that the X-ray period is equal to $561.64\pm0.56$~s. Because the last two periods correspond to each other, this confirms the IP nature of V1033~Cas. In addition, the detection of the X-ray period suggests that this period is most probably the spin period of the white dwarf, but not the sideband period (e.g., \citealt{hellier01}).  However, both detected periods have low precision and are not suitable to investigate changes of the spin period of the white dwarf in the future. Indeed, if we suppose an oscillation ephemeris with the spin period measured by \citeauthor{ bonnet07}, then the formal validity of this ephemeris will be only 6 days.  To measure the spin period with high precision, to obtain a spin oscillation ephemeris with a long validity and to learn other properties of V1033~Cas, we performed extensive photometric observations within 34 nights. These observations have a total duration of 134 hours and cover 11 months. In this paper, we present the results obtained from these observations.

\begin{table}[t]
{\small 
\caption{Journal of the observations.}
\label{journal}
\begin{tabular}{@{}l c c}
\hline
\noalign{\smallskip}
Date  &  BJD$_{\rm TDB}$ start & Length \\
(UT) & (-245\,0000) & (h) \\
\hline
2017 Jan. 30   & 7784.194728  &  8.0  \\
2017 Jan. 31   & 7785.092285 & 9.6 \\
2017 Feb. 16   & 7801.134348 & 4.0   \\
2017 Feb. 17   & 7802.129536 & 1.3  \\
2017 Mar. 16   & 7829.168016 & 2.6  \\
2017 Mar. 19   & 7832.173661 & 5.8  \\
2017 Mar. 20   & 7833.178435 & 6.8  \\
2017 Mar. 21    & 7834.182622 & 2.9 \\
2017 Mar. 22    & 7835.178989 & 7.6 \\
2017 Mar. 24    & 7837.281417 & 1.5  \\
2017 Mar. 27   & 7840.189258 & 2.7  \\
2017 Mar. 31    & 7844.202522 & 2.7  \\
2017 Apr. 1    & 7845.290858 & 4.3 \\
2017 Apr. 25    & 7869.286424 & 2.9 \\
2017 Apr. 26    & 7870.254826 & 2.3 \\
2017 Apr. 30    &  7874.275348 & 3.1 \\
2017 May 1   & 7875.306878  & 2.2 \\
2017 Aug. 24    & 7990.306557 & 3.3 \\
2017 Aug. 26    & 7992.349511 & 2.5 \\
2017 Aug. 28    & 7994.313131 & 3.6 \\
2017 Sep. 12    & 8009.221367 & 1.1 \\
2017 Sep. 15    & 8012.191499 & 7.0 \\
2017 Sep. 16  & 8013.405991 & 2.0 \\
2017 Sep. 19    & 8016.404643 & 2.1 \\
2017 Sep. 22    & 8019.161590 & 1.3 \\
2017 Oct. 10    & 8037.124348 & 1.6 \\
2017 Oct. 14    & 8041.121558 & 7.3 \\
2017 Nov. 21 & 8079.382237 & 1.9 \\
2017 Nov. 22   & 8080.206350 & 3.1 \\
2017 Nov. 24    & 8082.343900   & 6.4 \\
2017 Nov. 26    & 8084.219557 & 8.8 \\
2017 Nov. 27    & 8085.399986 & 1.7 \\
2017 Dec. 12    & 8100.056677 & 5.2 \\
2017 Dec. 13    & 8101.063948 & 4.5 \\
\hline
\end{tabular} }
\end{table}


\section{Observations} \label{observations}

For observations of variable stars, we apply a multi-channel pulse-counting photometer with photomultiplier tubes. The photometer is attached to the 70-cm telescope at Kourovka observatory, Ural Federal University. This photometer makes it possible to measure the brightness of two stars and the sky background simultaneously and almost continuously. Short and rare interruptions appear when we measure the sky background in all three channels simultaneously. This is necessary to define differences in the sky background caused by differences in the size of the diaphragm and other instrumental and atmospheric effects. To maintain the centring two stars in the photometer diaphragms, we first carried out observations by manually guiding with an eyepiece and a guiding star. Later we included the CCD guiding system in the photometer. Because of this guiding system, the photometer and the telescope can work automatically under computer control. Precise automatic guiding improves the accuracy of brightness measurements. In addition, it allows us to use lesser diaphragms, reducing the effect of the sky background. The design of the photometer and its noise analysis are described in \citet{kozhevnikoviz}.

Our photometer provides high-quality photometric data even under unfavourable atmospheric conditions \citep[e.g.,][]{kozhevnikov02}. For a long time, however, we could not observe stars fainter than 15~mag, which are invisible by eye (a 70-cm telescope). A few years ago we understood how to make the relative centring of two stars in the photometer diaphragms, one of which is invisible by eye, using the coordinates of the invisible star, the coordinates of the nearby reference star and computer-controlled step motors of the telescope. Using this method of centring, we can observe very faint stars up to 20~mag \citep[e.g.,][]{kozhevnikov18}. For such faint stars, continuous measurements of the sky background in the third channel, which are possible in our photometer, are extremely important because the counts of the sky background in the stellar diaphragm are much larger than the star counts. Therefore, photometric observations of very faint stars, which are invisible by eye, seem impossible without a sky background channel.

The photometric observations of V1033~Cas were performed in 2017 January--December over 34 nights with a total duration of 134~h. The data were obtained in white light (approximately 3000--8000~\AA). The time resolution was 4 s. For V1033~Cas and the comparison star, we used 16-arcsec diaphragms. To reduce the photon noise caused by the sky background, we measured the sky background through a diaphragm of 30~arcsec. The comparison star is USNO-A2.0 1500-00409838. It has $B=15.0$~mag and $B-R=0.6$~mag. We chose this star among neighbouring stars because the colour index of V1033~Cas is similar to the colour index of this star.  According to the USNO-A2.0 catalogue, V1033~Cas has $B=16.6$~mag and $B-R=0.3$~mag.  The similarity of the colour indexes reduces the influence of differential extinction. 
\begin{figure}[t]
\includegraphics[width=84mm]{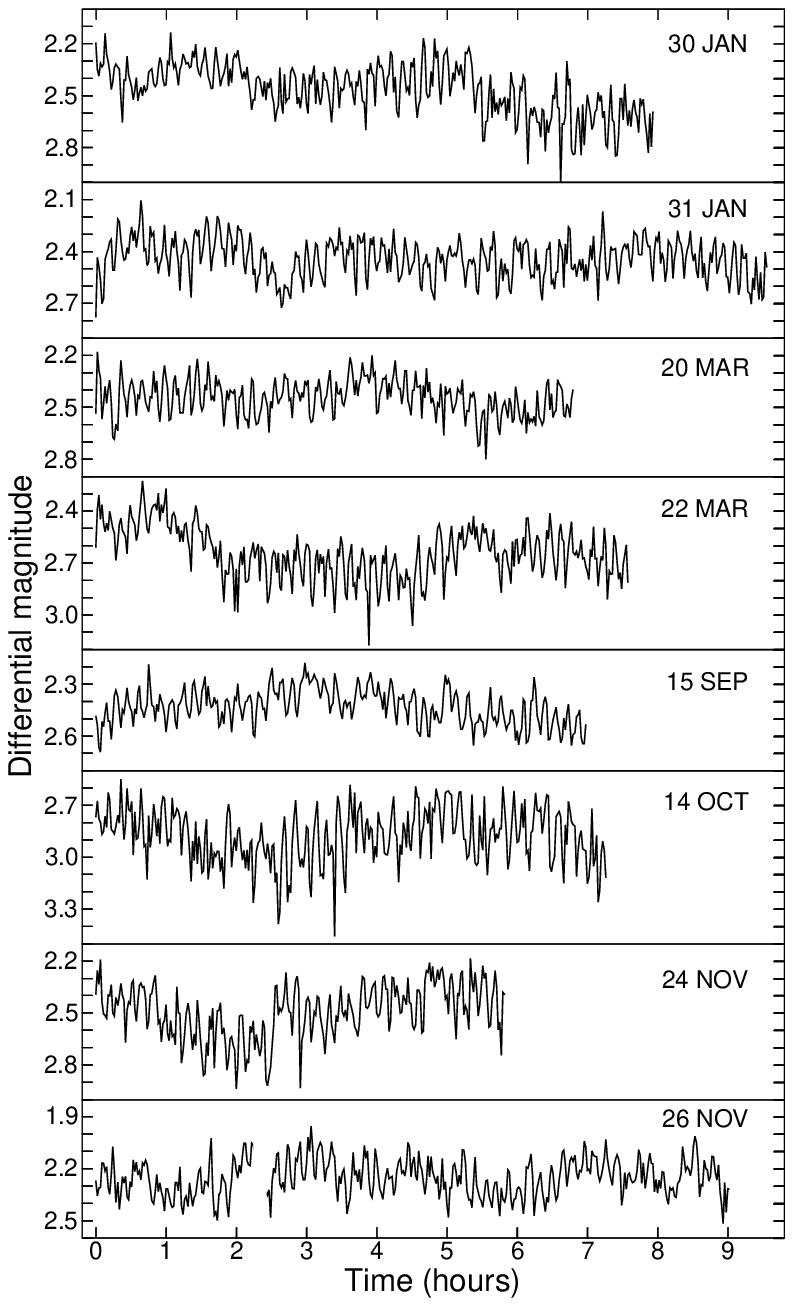}
\caption{Longest differential light curves of V1033~Cas. The short-period oscillation is clearly visible in all light curves}
\label{figure1}
\end{figure}

A journal of the observations is presented in Table~\ref{journal}. This table contains BJD$_{\rm TDB}$, which is the  Barycentric Julian Date in the Barycentric Dynamical Time (TDB) standard. BJD$_{\rm TDB}$ is preferential because this time is uniform. We calculated BJD$_{\rm TDB}$ using the online-calculator (http://astroutils.astronomy.ohio-state.edu/time/) \citep{eastman10}. In addition, using the BARYCEN routine in the 'aitlib' IDL library of the University of T\"{u}bingen (http://astro.uni-tuebingen.de/software/idl/aitlib/), we calculated BJD$_{\rm UTC}$, the Barycentric Julian Date in the Coordinated Universal Time (UTC) standard. The difference between BJD$_{\rm TDB}$ and BJD$_{\rm UTC}$ was constant during our observations of V1033~Cas. BJD$_{\rm TDB}$ exceeded BJD$_{\rm UTC}$ by  69~s.

The obtained differential light curves have high photon noise due to the low brightness of V1033~Cas (roughly 17.5~mag) and the high sky background. Fig.~\ref{figure1} presents the longest differential light curves of V1033~Cas. To reduce the photon noise, the magnitudes in these light curves were previously averaged over 80-s time intervals. According to the pulse counts of the two stars and the sky background, the photon noise of these light curves (rms) is 0.04--0.08~mag. As seen in Fig.~\ref{figure1}, despite this large photon noise, the short-period oscillation with a period of 563~s, which is detected in X-rays \citep{anzolin09} and in optical light \citep{bonnet07} and corresponds to the spin period of the white dwarf, is clearly visible in all light curves due to the large oscillation amplitude. 
\begin{figure}[t]
\includegraphics[width=84mm]{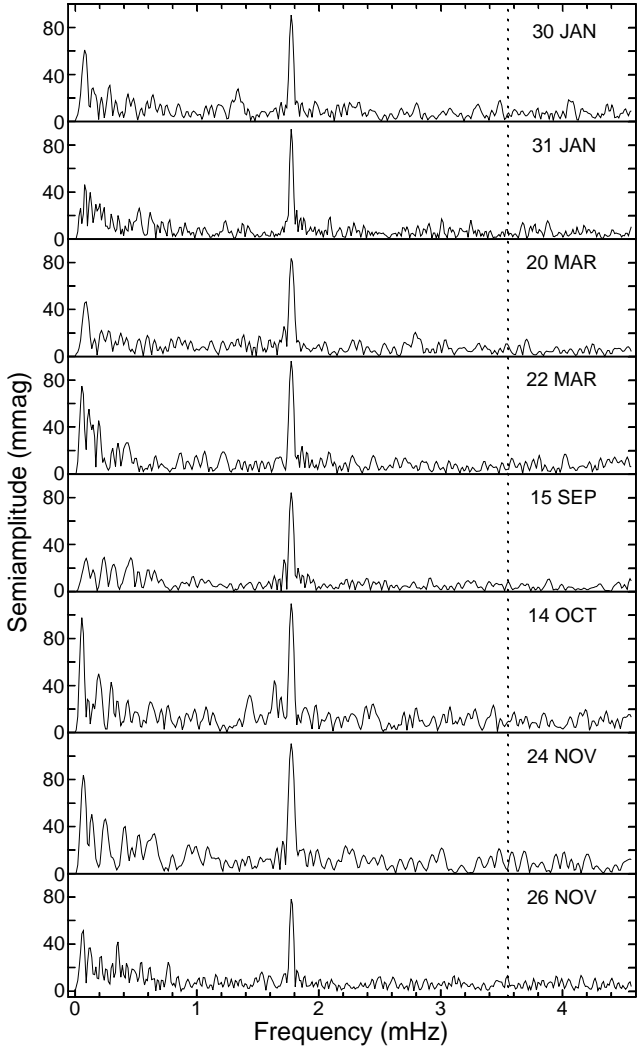}
\caption{Amplitude spectra of V1033~Cas. The prominent peak, which corresponds to the 563-s period, shows an almost constant height of about 90~mmag. The dotted line marks the first harmonic of the 563-s period}
\label{figure2}
\end{figure}

\section{Analysis and results}

In addition to the short-period oscillation, the longest differential light curves of V1033~Cas presented in Fig.~\ref{figure1} show obvious flickering, which is typical of all CVs. There are many complicated methods to investigate flickering properties in CVs such as power spectral density \citep{scaringi14}, wavelet transforms \citep{fritz98} etc. Here, to characterize the flickering power in V1033~Cas, we use the simplest method, namely, the flickering peak-to-peak amplitude. This allows us to compare the flickering power in V1033~Cas with the flickering power in several IPs, which we observed using the same technique.  The flickering power in V1033~Cas seems weak and similar to the flickering power we observed in most other IPs. From Fig.~\ref{figure1}, we estimate that the flickering peak-to-peak amplitude is equal to 0.2--0.3~mag. V709~Cas, V515~And and V647~Cyg revealed similar flickering peak-to-peak amplitudes of less than 0.4 mag \citep[see Figs.~1 in][]{kozhevnikov01, kozhevnikov12,  kozhevnikov14}. The flickering power in these IPs seem weak compared with the flickering power in V2069~Cyg, which showed a noticeably larger flickering peak-to-peak amplitude of 0.4--0.6~mag \citep[see Fig.~1 in][]{kozhevnikov17}.

Our multichannel photometer allows us to obtain evenly spaced data. Such data are advantageous compared with unevenly spaced data because they allow us to use the classical Fourier analysis. In addition, the Fourier analysis seems preferable for sinusoidal or smooth quasi-sinusoidal signals \citep[e.g.,][]{schwarzenberg98}. It is especially suitable for IPs because short-period oscillations of sinusoidal and quasi-sinusoidal shapes are typical of IPs in optical light. Using a fast Fourier transform (FFT) algorithm, we calculate individual amplitude spectra and power spectra of the data included in common time series. Before applying the FFT algorithm, we eliminate low-frequency trends from individual light curves by subtraction of a first-, second- or third order polynomial fit. This is a usual procedure in Fourier analysis and does not affect high frequencies.

Despite the large photon noise, the spin oscillation in V1033~Cas is clearly visible in the light curves (Fig.~\ref{figure1}). The obvious reason for the good visibility of the spin oscillation is the large amplitude.  Indeed, as seen in Fig.~\ref{figure2}, which shows the amplitude spectra of V1033~Cas, the spin oscillation  semi-amplitude is  approximately 90~mmag.  In addition, Fig.~\ref{figure2} demonstrates that changes of the oscillation amplitude from night to night are not significant.  At frequencies below 1~mHz, the amplitude spectra show many small peaks. These peaks are randomly distributed in frequency and are apparently caused by flickering.

\begin{figure}[t]
\includegraphics[width=84mm]{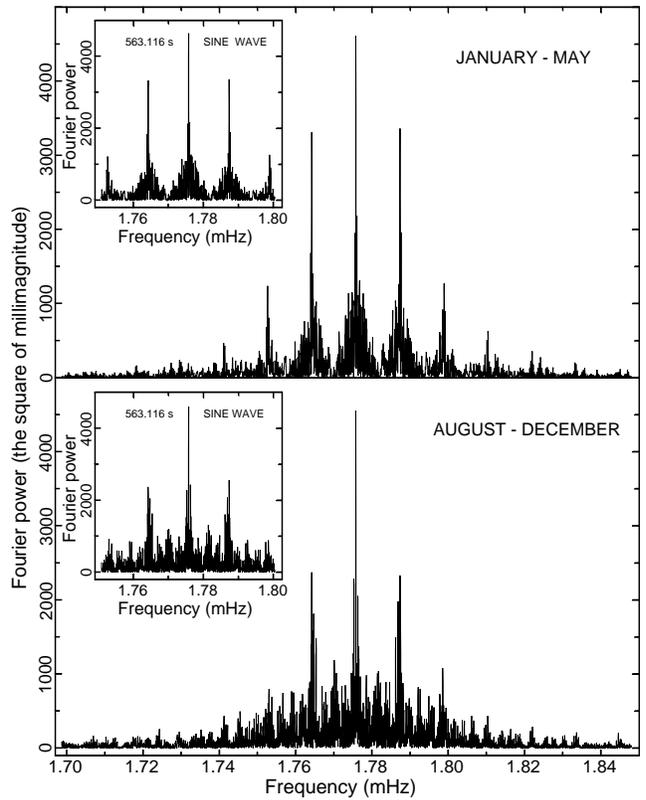}
\caption{Power spectra calculated for two groups of the data of V1033~Cas. They show the coherent oscillation with periods of $563.116\,83\pm0.000\,60$ (upper frame) and $563.116\,90\pm0.000\,50$~s (lower frame). Insets show the window functions derived from artificial time series}
\label{figure3}
\end{figure}

The individual amplitude spectra shown in Fig.~\ref{figure2} provide comprehensive information about the amplitude of the spin oscillation and its behaviour in time. Using these amplitude spectra, however, we cannot define the precise oscillation period because their frequency resolution is very low.  Therefore, we analysed the data included into common time series. As mentioned, we eliminated low-frequency trends from individual light curves by subtraction of a first-, second- or third order polynomial fit. After such elimination of trends, the nightly averages of the individual light curves are equal to zero. Therefore, the gaps due to daylight and poor weather were filled with zeros according to the time of observations. This is optimal because gaps filled with zeros do not introduce additional discontinuity to the data. In addition, zeros give no additions to the Fourier transform. Due to the large observation coverage, the power spectra of the common time series have a much higher frequency resolution. To avoid the aliasing problem and find out changes in the oscillation behaviour, we first divided the V1033~Cas light curves into two parts, namely, the data of January--May and the data of August--December. These parts of data are separated by a large gap from each other (see Table~\ref{journal}).  Fig.~\ref{figure3} shows the power spectra of two common time series containing data from these two parts. These power spectra reveal distinct principal peaks and one-day aliases corresponding to the spin oscillation. In addition, they show fine structures that coincide with the fine structures of the window functions. This proves the absence of the aliasing problem. In addition, this is a strong indication that the spin oscillation is coherent both during January--May and during August--December.

To find the precise values of the spin period, we used a Gaussian function fitted to the upper parts of the principal peaks. The errors are defined according to the method of \citet{schwarzenberg91}.  This method considers both frequency resolution and noise. It was tested in our previous works, where we made sure that errors according to \citeauthor{schwarzenberg91} are true rms errors (e.g., \citealt{kozhevnikov12}). The values of the spin period found from the power spectra presented in Fig.~\ref{figure3} are $563.116\,83\pm0.000\,60 $ and $563.116\,90\pm0.000\,50 $~s in January--May and in August--December, respectively. These values correspond to each other, because they differ only by $0.1\sigma$. Although in each of these two cases we did not use all observational coverage, nevertheless, the precision of the oscillation periods was high. This high precision is achieved due to the low noise level, which is caused by the large oscillation amplitude. The spin oscillation semi-amplitudes found from the power spectra shown in Fig.~\ref{figure3} are equal to 96 and 95~mmag  in January--May and in August--December, respectively. These semi-amplitudes are compatible with the heights of the peaks visible in the individual amplitude spectra (Fig.~\ref{figure2}).

\begin{figure}[t]
\includegraphics[width=84mm]{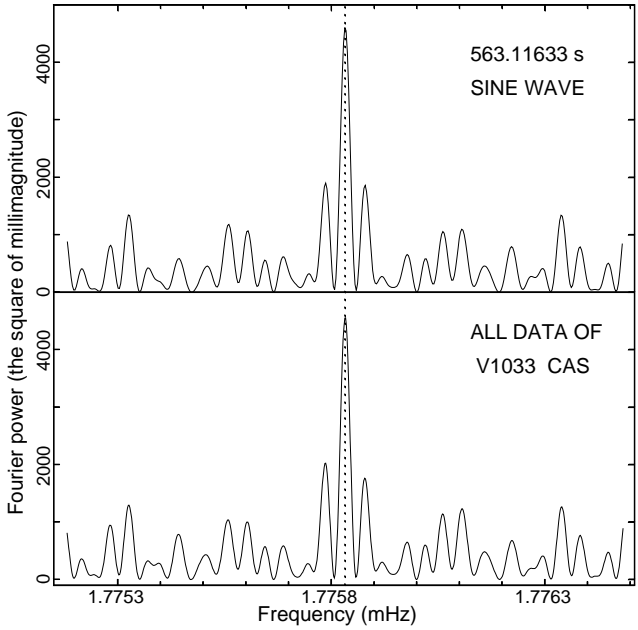}
\caption{Power spectrum segment around the spin oscillation calculated for all V1033~Cas data. It reveals a period of $563.116\,33\pm 0.000\,10 $~s. The upper frame shows the window function. Note the close resemblance of the window function and the power spectrum}
\label{figure4}
\end{figure}

Making sure that the two parts of the data provide distinct power spectra without the aliasing problem, we included all 34 observational nights of V1033~Cas into the common time series. Fig.~\ref{figure4} presents the power spectrum segment of this common time series around the spin oscillation. Obviously, this power spectrum gives the most precise spin period due to the highest frequency resolution and due to the lowest noise level. As seen, the fine structure of this power spectrum is very similar to the fine structure of the window function. In addition, two peaks near the principal peak, which are caused by the large gap between the two parts of data, have much lower heights than the height of the principal peak. Obviously, this eliminates the aliasing problem. The close similarity of this power spectrum and the window function proves the complete coherence of the spin oscillation during the eleven months of our observations. The period and semi-amplitude of the spin oscillation found from the power spectrum presented in Fig.~\ref{figure4} are equal to $563.116\,33\pm 0.000\,10$~s and 95~mmag, respectively. They are compatible with the periods and semi-amplitudes determined from the power spectra of the two parts of data.

\begin{table}
\small
\caption{The values and precisions of the spin period}
\label{table2}
\begin{tabular}{@{}l l l l l l}
\hline
\noalign{\smallskip}
Time                 & Semi-amp.      & Period & Error & Dev. \\
span                & (mmag)            & (s)       & $\sigma$ (s)  &     &    \\
\hline
Jan.--May        & $95.5\pm2.0$    & 563.11683 &   0.00060    & $0.8\sigma$    \\
Aug.--Dec.       & $95.4\pm2.4$   & 563.11690 &   0.00050    & $1.1\sigma$    \\
Total                & $95.5\pm1.3$   & 563.11633 &  0.00010     & -- \\
\hline
\end{tabular}
\end{table}

Summarized information about the periods and the amplitudes of the spin oscillation is presented in Table~\ref{table2}. Precise semi-amplitudes and their rms errors are found from a sine wave fitted to the folded light curves. These semi-amplitudes are close to the semi-amplitudes found from the power spectra. The spin oscillation semi-amplitudes are surprisingly constant in the two parts of data. In the fourth column, we give the rms errors of the periods obtained by the method of \citet{schwarzenberg91}. The spin period error found from all data is much lower than other errors. We found deviations of the other periods and expressed them in units of their rms errors. This is shown in the fifth column. As seen, these deviations are not excessively small and obey a rule of $3\sigma$.  As in our previous works \citep{kozhevnikov12, kozhevnikov14, kozhevnikov17}, this confirms that the errors calculated according to \cite{schwarzenberg91} are true rms errors.

\begin{figure}[t]
\includegraphics[width=84mm]{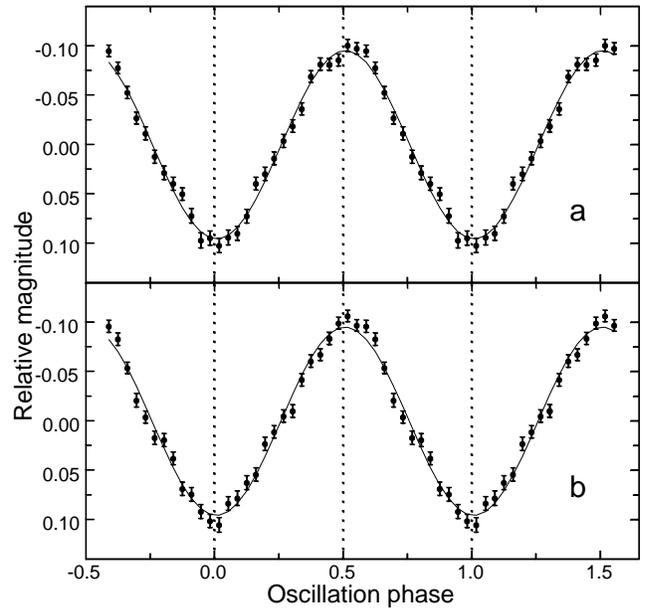}
\caption{Pulse profiles of the spin oscillation obtained from the data of January--May (a) and from the data of August--December (b). The continuous curves are sine waves fitted to the folded light curves. Note their very close semi-amplitudes: $95.5\pm2.0$~mmag in frame (a) and $95.4\pm2.4$~mmag in frame (b)}
\label{figure5}
\end{figure}

To investigate the spin pulse profiles, we folded the light curves of two parts of data with a period of $563.116\,33$~s. Fig.~\ref{figure5} presents the results. The continuous curves are sine waves fitted to the folded light curves. These two sine waves have remarkably close semi-amplitudes, namely, $95.5\pm2.0$ and $95.4\pm2.4$~mmag for the data of January--May and for the data of August--December, respectively. In addition, the pulse profiles obtained from the two parts of data are also similar to each other.  As seen, despite the small rms errors, the points in the folded light curves deviate only slightly from the fitted sine waves. Deviations do not exceed 2$\sigma$. Although some consecutive points show one-sided deviations, these deviations are also statistically insignificant. Indeed, if we combine the points of the folded light curves into groups of three points and average them, then the deviations from the sine waves will not exceed 3$\sigma$. Thus, the spin pulse profiles are sinusoidal with high accuracy. This is compatible with the absence of the first harmonic of the spin oscillation in the amplitude spectra (Fig.~\ref{figure2}).

\begin{figure}[t]
\includegraphics[width=84mm]{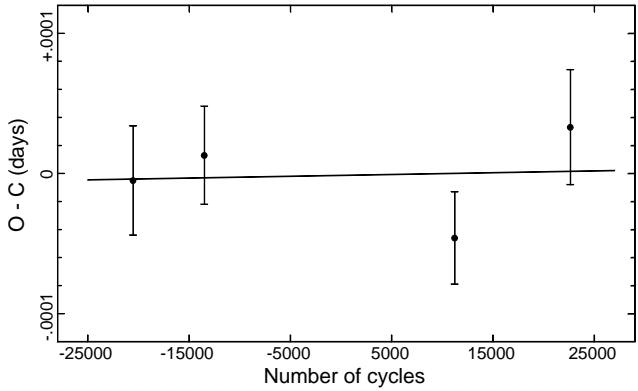}
\caption{(O--C) diagram for all V1033~Cas data, which are subdivided into four groups}
\label{figure6}
\end{figure}

Using the spin period measured with high precision, we can derive an oscillation ephemeris with a long validity. In addition to the spin period, we must determine the phase of the oscillation. Obviously, because of photon noise and flickering, direct measurements of oscillation phases can be obtained only with low precision.  Therefore, we obtained the oscillation phase from the folding of all data. Because the spin pulse profile is sinusoidal, to find the time of spin pulse maximum, it is convenient to use a sine wave fitted to the folded light curve. The resulting time of spin pulse maximum was referred to the middle of the observations. In addition, we used the data divided into four groups for verification. The times of spin pulse maximum for these groups were also referred to the middle of the corresponding data groups. These times are presented in Table~\ref{table3}. Using the main time of spin pulse maximum, we obtained the following ephemeris:

{\small
\begin{equation}
BJD_{\rm TDB} (\rm max)= 245\,7942.727\,207(21)+0.006\,517\,5501(12) {\it E}.
\label{ephemeris}	
\end{equation} }

\begin{table}
\scriptsize
\caption{Verification of the spin ephemeris}
\label{table3}
\begin{tabular}{@{}l l l c}
\hline
\noalign{\smallskip}
Time         & BJD$_{\rm TDB}$(max)             & N. of    & O--C$\times 10^{3}$     \\
span        & (-245\,0000)                               & cycles  & (days)    \\
\noalign{\smallskip}
\hline
Jan. 30--Mar.  20     & 7808.837170(32)    & --20543     &  --0.005(39)   \\ 
Mar. 21--May 1        & 7854.798952(28)    & --13491     &  +0.013(35)   \\
Aug. 24--Oct. 14      & 8015.873626(26)   & +11223     &  --0.046(33)     \\
Nov. 21--Dec. 13     & 8090.323680(35)     & +22646    &  +0.033(41)   \\
\hline
\end{tabular}
\end{table}

To verify the oscillation ephemeris, we calculated $(O - C)$ values and numbers of cycles for the four groups of data (Table~\ref{table3}). The $(O - C)$ diagram is presented in Fig.~\ref{figure6}. It reveals no significant slope and displacement along the vertical axis. Indeed, the calculated $(O - C)$ values obey the relation: ${\rm (O - C)}= - 0.000\,008(67) - 0.000\,000\,0004(33) {\it E}$. All quantities in this relation are less than their rms errors. Therefore, the ephemeris does not require any correction. According to the rms error of the spin period, the formal validity time of this ephemeris is 100 years (a confidence level of $1\sigma$).  

To analyse the short-period oscillations, we removed the low-frequency trends from the individual light curves by subtraction of a first-, second- or third order polynomial fit. To search for the orbital variability of V1033~Cas, such a procedure cannot be used because many of our individual light curves are much shorter than the orbital period of V1033~Cas. According to the radial velocity measurements by \citet{bonnet07}, the orbital period of V1033~Cas is $4.033\pm0.005$~h. Obviously, The low frequencies corresponding to the orbital variability of V1033~Cas will be removed from the short light curves due to  this removal of trends. In addition, by performing numerical experiments with artificial time series, we made sure that we also could not remove the nightly averages from the  individual light curves because such removal would significantly reduce the orbital variability of V1033~Cas. Therefore, we removed only two common averages from two groups of data.  
\begin{figure}[t]
\includegraphics[width=84mm]{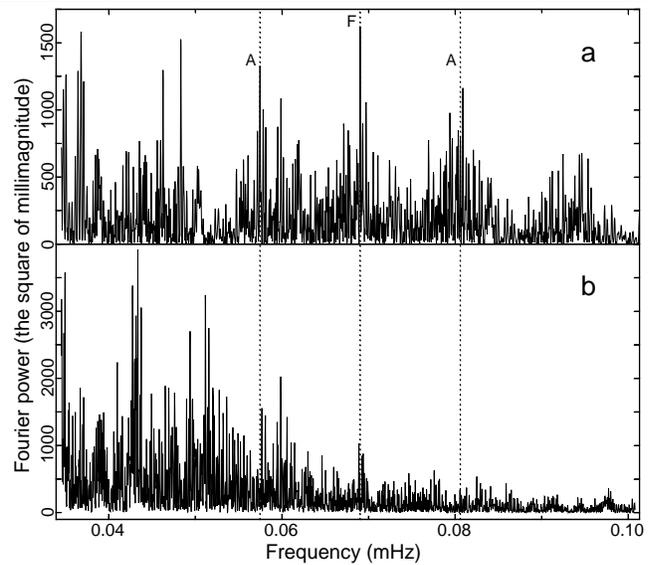}
\caption{Power spectra of V1033~Cas in the frequency range around the expected orbital variability. The data of January--May (frame a) reveal the prominent principal peak which coincides with the orbital period found by \citeauthor{bonnet07} (labelled F). The left-hand one-day alias is also prominent (labelled A). In addition, the small right-hand one-day alias is perceptible (labelled A). In contrast, the data of August--December (frame b) reveal no peaks corresponding to the orbital period. Note that the vertical scale in frame b is two times less}   
\label{figure7}
\end{figure}

The resulting power spectra are presented in Fig.~\ref{figure7}. As seen, at least in the data of January--May, the power spectrum shows the prominent principal peak and its left-hand one-day alias, which correspond to the orbital period found by \citeauthor{bonnet07} (Fig.~\ref{figure7}a). From this power spectrum, the orbital period and the semi-amplitude are equal to $4.0243\pm0.0028$~h and 46~mmag, respectively.  Here, the error of the orbital period is equal to the half-width of the principal peak at half maximum. In the data of August--December, the power spectrum does not show signs of the orbital variability of V1033~Cas (Fig.~\ref{figure7}b). The data of August--December are not favourable to detect the orbital variability of V1033~Cas because of shorter lengths of individual light curves and larger brightness changes from night to night (see Fig.~\ref{figure1}), which create additional noise in the power spectrum.

\begin{figure}[t]
\includegraphics[width=84mm]{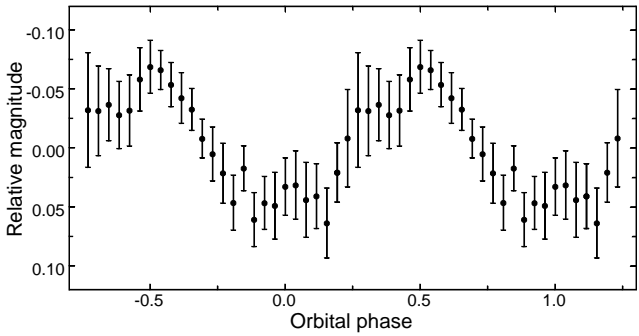}
\caption{Orbital pulse profile of V1033~Cas obtained from the data of January--May. The pulse profile is quasi-sinusoidal with slightly wider minima compared with maxima and shows a semi-amplitude of $55\pm4$~mmag}

\label{figure8}
\end{figure}

To find the orbital pulse profile, we folded the data of January--May, in which the orbital variability was detected, with a period of 4.0243~h. To avoid distortions of the orbital pulse profile by the spin oscillation, the data were previously averaged over 564-s time intervals. As seen in Fig.~\ref{figure8}, the orbital pulse profile of V1033~Cas is quasi-sinusoidal  with slightly wider minima compared with maxima. The orbital pulse profile shows a semi-amplitude of $55\pm4$~mmag. The semi-amplitude and its rms error are found from a sine wave fitted to the folded light curve. This semi-amplitude is compatible with the semi-amplitude derived from the power spectrum shown in Fig.~\ref{figure7}a.

\begin{figure}[t]

\includegraphics[width=84mm]{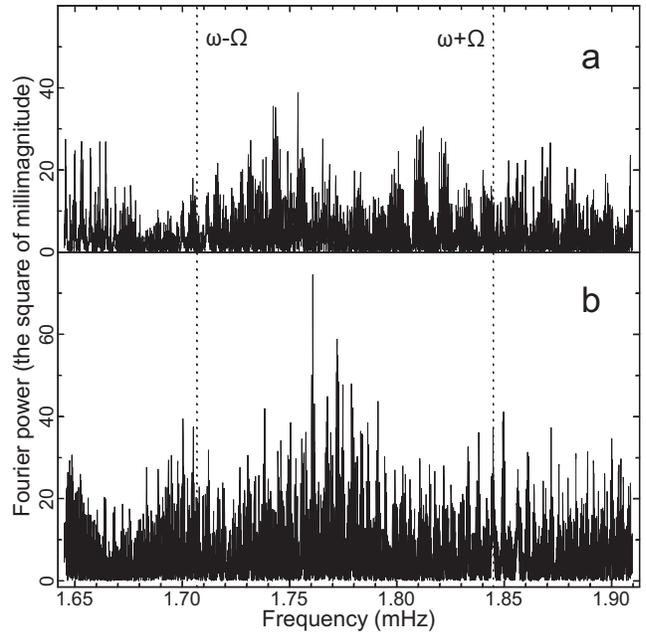}
\caption{Power spectra of two parts of the V1033~Cas data, from which the spin oscillation is excluded. Both the power spectrum of the data of January--May (frame a) and the power spectrum of the data of August--December (frame b) do not show coherent oscillations. The expected frequencies of the negative and positive orbital sidebands are labelled $\omega - \Omega$ and $\omega + \Omega$, respectively}

\label{figure9}

\end{figure}

In many cases, the spin oscillations in IPs are accompanied with oscillations in the orbital sidebands, which can be both negative and positive, i.e., $\omega - \Omega$ and $\omega + \Omega$.  These oscillations have the periods close to the spin period. Therefore, when such oscillations have small amplitudes, they can be hidden in the complex structure of the window function created by the spin oscillation. To find orbital sidebands, we applied the commonly used method of extraction of the main oscillation from the data. Fig.~\ref{figure9} presents the power spectra of two parts of data, from which the spin oscillation is excluded. As seen, despite the large semi-amplitude of the spin oscillation, this oscillation is completely excluded. This, however, did not lead to the detection of any sideband oscillations. Obviously, the undetected sideband oscillations should have small amplitudes, which do not exceed the amplitudes of the noise peaks. The maximum semi-amplitudes of the noise peaks are 9 and 12~mmag in the data of January-May and in the data of August--December, respectively. Thus, the sideband semi-amplitudes in V1033~Cas should not exceed roughly 10~mmag and should be at least ten times less than the semi-amplitude of the spin oscillation.

\section{Discussion}
We performed photometric observations of V1033~Cas with a total duration of 134~h that cover eleven months in 2017. The comprehensive Fourier analysis of these data allowed us to measure the spin period of the white dwarf in V1033~Cas with high precision, $P_{\rm spin}=563.116\,33\pm0.000\,10$~s.  This high precision was achieved due to the large observational coverage and due to the low noise level. Although the spin period of the white dwarf in V1033~Cas was measured previously both in X-rays \citep[$561.64\pm0.56$~s,][]{anzolin09} and in optical light \citep[$563.53\pm0.62$~s,][]{bonnet07}, their precision was much lower because of insufficient observational coverage. Indeed, \citeauthor{anzolin09} observed V1033~Cas within 6.9~h, and \citeauthor{bonnet07} observed V1033~Cas within 5.3~h. Such short observations have a low frequency resolution and cannot provide high precision of the oscillation period. In addition, such short observations have a large noise level. However, according to the rms errors, the spin period measured by us is compatible with the spin periods measured by \citeauthor{bonnet07} and by \citeauthor{anzolin09}

During our observations, the spin oscillation semi-amplitude in V1033~Cas was stable and was equal to $95.5\pm1.3$~mmag. This semi-amplitude is large compared with the semi-amplitudes of the spin oscillations in other IPs. Indeed, as seen in Table 1 in \citet{patterson94}, with the exception of VZ Pyx which was recently recognized as a dwarf nova with superhumps, only two IPs show larger semi-amplitudes. They are FO~Aqr (0.18~mag) and EX~Hya (0.24~mag). These IPs, however, have much larger spin periods. Therefore, according to the period--amplitude relation \citep[see Fig.~12 in][]{patterson94}, they must have large amplitudes.  The spin oscillation in V1033~Cas was the strongest oscillation compared with the oscillations in the 7 later discovered IPs, which we observed by using the same technique \citep{kozhevnikov01, kozhevnikov10, kozhevnikov12, kozhevnikov14, kozhevnikov15, kozhevnikov16, kozhevnikov17}. Indeed, among these 7 IPs, only MU Cam revealed a bit lesser semi-amplitude of 90~mmag. The semi-amplitudes seen in other 6 IPs were in the range 0.005-0.041~mag. Although MU~Cam shows almost the same semi-amplitude as V1033~Cas, this IP has a noticeably larger spin period (1187~s). To find other IPs with large spin oscillation amplitudes and with short spin periods, which were discovered after the work by \citeauthor{patterson94}, we used the IP homepage of Mukai (https://asd.gsfc.nasa.gov/Koji.Mukai/iphome/iphome.html) and references therein. The catalogue of IPs presented in this homepage was updated at the end of 2014. We identified 17 ironclad and confirmed IPs with spin periods of less than 600~s and with known optical amplitudes. We learned that V418~Gem shows the largest semi-amplitude, namely, 150~mmag \citep{gansicke05}. However, this is the semi-amplitude of the first harmonic of the spin oscillation in V418~Gem.  If the true spin oscillations are considered, then V1033~Cas will show the larger semi-amplitude. The second largest semi-amplitude (65~mmag) belongs to HT~Cam \citep{demartino05}. In addition, we considered 8 recently discovered IPs \citep{halpern18}, where 6 of them were confirmed due to optical oscillations. We learned that, among them, the largest semi-amplitude was observed in 2PBC~J1911.4+1412 (0.25~mag). However, the spin period of this IP (747~s) is noticeably larger than the spin period in V1033~Cas. Among IPs, which were described in \citeauthor{halpern18}, only 1SWXRT~J230642.7+550817 has a lesser spin period of 464~s. Its oscillation semi-amplitude (50~mmag) is also noticeably lesser than the semi-amplitude we observed in V1033~Cas. Thus, the oscillation semi-amplitude in V1033~Cas seems very large among IPs with similar and lesser spin periods. Due to the low relative noise level caused by flickering at higher frequencies, the large oscillation amplitude in V1033~Cas is favourable to obtain the high precision of the spin period and to investigate changes of the spin period using $O - C$.  11 years ago, however, the semi-amplitude of the spin oscillation in V1033~Cas was two times less \citep[see][]{bonnet07}.

Although most of our individual light curves of V1033~Cas are short compared with the orbital period detected by \citet{bonnet07} from radial velocity measurements, we were nevertheless able to find the corresponding photometric variations with a period of $4.0243\pm0.0028$~h and with a semi-amplitude of $55\pm4$~mmag. The orbital variability of V1033~Cas was also noticed by \citeauthor{bonnet07} This allowed us to roughly estimate the orbital inclination of V1033~Cas. Because our light curves folded with the orbital period showed no eclipses (Fig.~\ref{figure8}), the orbital inclination should be less than $60^\circ$ \citep[e.g.,][]{ladous94}. Another reason for orbital variability is the presence of orbital humps in the light curves, which are caused by a hot spot in the accretion disc. They are noticeable if the orbital inclination is larger than $50^\circ$ \citep{ladous94}. If the orbital inclination is less than $50^\circ$, the orbital variations of the novalike variables, to which most IPs belong, are difficult to detect because their accretion discs are much brighter than their stellar components \citep[e.g.,][]{bonnet01}. Therefore, ellipsoidal modulations, reflection effects, etc. should be inconspicuous. Thus, the orbital inclination of V1033~Cas is probably $50^\circ - 60^\circ$. 

The estimated orbital inclination of V1033~Cas roughly corresponds to its large spin oscillation amplitude. As noted by \citet{hellier91}, the spin oscillation amplitude is greatest when the system is viewed edge-on. As mentioned, MU~Cam shows almost the same spin oscillation semi-amplitude as V1033~Cas. In addition, MU~Cam shows orbital variations with a similar semi-amplitude of 42--59~mmag \citep[see Fig.~5 in][]{kozhevnikov16}. The similar amplitudes of orbital variations in V1033~Cas and in MU~Cam can mean that the inclinations of these two systems are also similar. This explains the similar spin oscillations amplitudes in these two IPs. In addition, according to \citet{hellier91}, the large spin oscillation amplitudes in these two systems suggest that they are noticeably inclined.

During our observations, the spin pulse profile was constant and sinusoidal with high accuracy (Fig.~\ref{figure5}). This is compatible with the sinusoidal pulse profile that was observed by \cite{bonnet07} 11 years ago. This very sinusoidal pulse profile is difficult to explain. For most IPs that do not show eclipses and do not rotate very fast, the accretion curtain model can explain optical spin pulses.  According to this model, optical emission originates from two accretion curtains located between the inner disc and the white dwarf \citep{hellier95}. Because these two accretion curtains can act in phase, the two accretion poles of the white dwarf can create a single-peaked roughly sinusoidal pulse profile. However, it seems strange how two accretion curtains viewed at different angles can produce a strictly sinusoidal pulse profile. This strictly sinusoidal pulse profile suggests that we see optical emission from one accretion curtain formed by one accretion pole.

Two-pole disc-fed accretion is considered to be the normal mode of behaviour in IPs because both magnetic poles of the white dwarf are on equal conditions with respect to the accretion disc \citep[e.g.,][]{warner95}. Therefore, we cannot explain the strictly sinusoidal pulse profile by supposing one-pole accretion. In addition, one of the two accretion curtains in an ordinary IP cannot be continuously hidden by a white dwarf because the length of accretion curtains is 4 -- 12 times larger than the white dwarf radius \citep[e.g.,][]{ferrario93} and because accretion curtains are extended over angles greater than $100^\circ$ \citep[e.g.,][]{hellier99}.  However, \citet{hellier99} performed Doppler tomography of accretion curtains in seven IPs and found that three of them show only the upper accretion curtains. These are high-inclination systems ($>60^\circ$), which show line emission only from the outer parts of the accretion curtains. Hence, the outer regions of the lower accretion curtain in such IPs can be hidden due to the obscuration of the lower accretion curtain by the inner edge of the accretion disc and the obscuration by the upper accretion curtain \citep{hellier99}. Although our estimate of the V1033~Cas inclination is somewhat less ($50^\circ - 60^\circ$), it can be sufficient, and the lower accretion curtain can be obscured by the disc and can be obscured by the upper accretion curtain. Thus, because the lower accretion curtain in V1033~Cas can be mostly invisible, we can explain the strictly sinusoidal spin pulse profile as if it was created by a single accretion curtain.

Despite the large amount of our observations, we did not find oscillations in the orbital sidebands. This agrees with the previous observations by \citet{bonnet07} and \citet{anzolin09}, who also did not find sidebands. As seen in Fig.\ref{figure9}, the semi-amplitudes of the undetected sideband oscillations do not exceed 10~mmag  and are at least ten times less than the spin oscillation semi-amplitude. Such weak signals seem undetectable in slowly spinning IPs.  Indeed, as seen in Table 1 in \cite{patterson94}, only two IPs show slightly lesser oscillation semi-amplitudes. These are rapidly spinning AE~Aqr and V533~Her with periods of 33.1 and 63.6~s and with semi-amplitudes of 5 and 7~mmag, respectively.  Hence, such weak signals may be below the detection threshold in slowly spinning IPs. Among the known IPs, there is a fairly distinct division between those that have a strong spin oscillation and those that have a strong sideband oscillation. Nonetheless, in many IPs, the alternative weak oscillation was detected \citep{warner95}. In addition, the stable optical oscillation and the presence of one or more orbital sidebands, even without any X-ray detection at all, provide strong evidence for inclusion in the IP class \citep{kuulkers06}. Therefore, the absence of detectable orbital sidebands in V1033~Cas at a semi-amplitude level of 10~mmag is unusual and difficult to explain.

Assuming a pure disk accretion, as indicated by the absence of orbital sidebands in X-rays, \citet{anzolin09} suggested that the white dwarf in V1033~Cas is not strongly magnetized and is not spinning at equilibrium. If so, the spin period of the white dwarf in V1033~Cas should change over large time spans. Using our precise oscillation ephemeris~\ref{ephemeris} and times of spin pulse maximum that were obtained earlier, we can verify the stability of the spin period.  Unfortunately, in the past, only the single time of spin pulse maximum was obtained in optical light by \citet{bonnet07}. Therefore, we also used the time of spin pulse maximum obtained in X-rays by \citet{anzolin09}. As noted by \citeauthor{anzolin09}, the time of spin pulse maximum in X-rays coincided with the time of spin pulse maximum in optical light. Unfortunately, in the past, other times of spin pulse maximum were not obtained. The times of spin pulse maximum, which were measured by \citeauthor{bonnet07} and \citeauthor{anzolin09}, were expressed in HJD$_{\rm UTC}$. We converted these times in BJD$_{\rm TDB}$. Using our ephemeris~\ref{ephemeris} we calculated $(O - C)$. Next, we calculated ${\rm d}P/{\rm d}t$ using the following formula \citep{Breger98}:

\begin{equation}
{\rm (O - C)} = 0.5 \, \frac{1}{p} \, \frac{{\rm d}P}{{\rm d}t} \, t^2.
\label{breger}	
\end{equation}

\begin{table}
\scriptsize
\caption{Stability of the spin period}
\label{table4}
\begin{tabular}{@{}l l l c}
\hline
\noalign{\smallskip}
BJD$_{\rm TDB}$(max)  & N. of          & O -- C         &    ${\rm d}P/{\rm d}t$     \\
        (-245\,0000)             & cycles       & (phases)       &                                                  \\
\noalign{\smallskip}
\hline
3975.51859(7)               & --608696     &  $-0.298\pm0.113$    &    $-(1.6\pm0.6)\times10^{-12}$    \\ 
4291.86747(20)            & --560158      &  $-0.293\pm0.108$    &    $-(1.9\pm0.7)\times10^{-12}$    \\
\hline
\end{tabular}
\end{table}

The results are shown in Table~\ref{table4}. Here, the upper line is calculated according to the time of spin pulse maximum measured by \citeauthor{bonnet07}, and the lower line is calculated according to the time of spin pulse maximum measured by \citeauthor{anzolin09} The fourth column gives the calculated ${\rm d}P/{\rm d}t$ and their rms errors. These two ${\rm d}P/{\rm d}t$, which are equal to $-(1.6\pm0.6)\times10^{-12}$ and   $-(1.9\pm0.7)\times10^{-12}$, match well each other because their difference is 3 times less than the summary rms error. In addition, these ${\rm d}P/{\rm d}t$  are only 2.7 times larger than their rms errors. Consequently, according to the triple rms error, the real ${\rm d}P/{\rm d}t$  must be less than 
$-4\times10^{-12}$. However, if we change the numbers of cycles to $\pm1$, $\pm2$ and $\pm3$, then these two ${\rm d}P/{\rm d}t$ will still be compatible with each other because  their difference will be less than $3\sigma$. In the last case, ${\rm d}P/{\rm d}t$ is roughly $\pm2\times10^{-11}$. With larger changes of the numbers of cycles, two ${\rm d}P/{\rm d}t$ become incompatible with each other because their difference become larger than $3\sigma$. Thus, the ${\rm d}P/{\rm d}t$  in V1033~Cas is most probably less than 
$-4\times10^{-12}$. This means that the spin period in V1033~Cas is very stable. With a low probability, the ${\rm d}P/{\rm d}t$  can reach $\pm2\times10^{-11}$.  However, even in this case, the spin period in V1033~Cas seems fairly stable because its possible change is less than the detection threshold of spin period changes in slowly spinning IPs (see Table~1 in \citealt{patterson94} and Table~1 in \citealt{warner96}).

The high stability of the spin period in V1033~Cas should be considered as a new result. This result seems very important because it contradict the suggestion that the white dwarf in V1033~Cas is not spinning at equilibrium \citep{anzolin09}. As a new result, this should be confirmed by future observations. This confirmation can be made from new photometric observations. Using new observations and our precise ephemeris~\ref{ephemeris}, one can calculate $(O - C)$ and analyse  their behaviour. However, as shown above, a small number of $(O - C)$ can give ambiguous results because of ambiguity in cycle numbers. Therefore, observations should be regular for ten years or more. Of course, such observations are difficult. Direct measurements of the spin period seem less difficult. By performing observations within 30--40 nights that cover a year, it is possible to achieve the same rms error of the spin period as we achieved in our observations, namely, 0.0001~s.  Then, if these observations are made ten years later, comparing two spin periods, one can reach ${\rm d}P/{\rm d}t$ of $1\times10^{-12}$ (a confidence level of $3\sigma$). This ${\rm d}P/{\rm d}t$ is 4 times less than the upper limit of ${\rm d}P/{\rm d}t$ we obtained from the analysis of $(O - C)$ described above. In addition, direct measurements of the spin period are devoid of ambiguity in cycle numbers.

\section{Conclusions}

We performed extensive photometric observations of V1033~Cas over 34 nights. The total duration of observations was 134~h. The observations covered 11 months. From the comprehensive analysis of these data, we obtained the following results: 
\begin{enumerate}
\item Due to the large observational coverage and the low noise level, we measured the spin period of the white dwarf with high precision. The spin period is equal to $563.116\,33\pm0.000\,10$~s. 
\item During our observations, the semi-amplitude of the spin oscillation was stable and was equal to $95.5\pm1.3$~mmag. This is a very large semi-amplitude of the spin oscillation among intermediate polars, which have similar and lesser spin periods. The large amplitude of the spin oscillation suggests that the system is noticeably inclined.
\item During our observations, the spin pulse profile was stable and sinusoidal with high accuracy. This may mean that we see the optical light produced by a single accretion curtain whereas the second accretion curtain may be obscured by the accretion disc.
\item Despite the large amount of observations, we could not detect sideband oscillations. The semi-amplitudes of the undetected sideband oscillations do not exceed 10~mmag and are at least ten times less than the semi-amplitude of the spin oscillation. The absence of sideband oscillations seems puzzling. 
\item Although our individual light curves are rather short compared with the orbital period of V1033~Cas, we were nevertheless able to detect orbital variations with a period of $4.0243\pm0.0028$~h and with a semi-amplitude of $55\pm4$~mmag.  This semi-amplitude seems large and suggests that the system is noticeably inclined. This is compatible with the large amplitude of the spin oscillation. 
\item The high precision of the spin period allowed us to obtain the oscillation ephemeris with a long validity of 100 years. Using this ephemeris and the times of spin pulse maximum obtained 10--11 years ago, we found that the spin period is very stable. ${\rm d}P/{\rm d}t$ is most probably less than  $-4\times10^{-12}$. This high stability contradicts the assumption that the white dwarf in V1033~Cas is not spinning at equilibrium.
\item Our precise ephemeris and our precise spin period can be used for future investigations of the stability of the spin period in V1033~Cas.
\end{enumerate}

\section*{Acknowledgments}

This work was supported in part by the Ministry of Education and Science (the basic part of the State assignment, RK No. AAAA-A17-117030310283-7) and by Program 211 of the Government of the Russian Federation (contract No. 02.A03.21.0006). This research has made use of the SIMBAD database, operated at CDS, Strasbourg, France. This research also made use of the NASA Astrophysics Data System (ADS).


\begin{thebibliography}{}


\bibitem[\protect\citeauthoryear{Anzolin et al.}{2009}]{anzolin09}
Anzolin,~G., de Martino,~D., Falanga,~M., Mukai,~K., Bonnet-Bidaud,~J.-M., et al.: \aap \, {\bf 501}, 1047 (2009) doi: 10.1051/0004-6361/200911816

\bibitem[\protect\citeauthoryear{Bikmaev et al.}{2006}]{bikmaev06}
Bikmaev,~I.~F., Revnivtsev,~ M.~G., Burenin,~R.~A., Sunyaev,~R.~A.: Astron. Lett. {\bf 32}, 588 (2006) 

\bibitem[\protect\citeauthoryear{Bonnet-Bidaud et al.}{2001}]{bonnet01}
Bonnet-Bidaud,~J.~ M, Mouchet,~ M., de~Martino,~ D., Matt,~G., Motch,~ C.: \aap \, {\bf 374}, 1003 (2001)

\bibitem[\protect\citeauthoryear{Bonnet-Bidaud et al.}{2007}]{bonnet07}
Bonnet-Bidaud,~J.~M., de Martino,~D., Falanga,~M., Mouchet,~M., Masetti,~N.: \aap \, {\bf 473}, 185 (2007) doi:10.1051/0004-6361:20077877 

\bibitem[\protect\citeauthoryear{Breger and Pamyatnykh}{1998}]{Breger98}
Breger,~M., Pamyatnykh,~A.~A.: \aap \, {\bf 332}, 958 (1998)

\bibitem[\protect\citeauthoryear{de~Martino et al.}{2005}]{demartino05}
de~Martino,~D., Matt,~G., Mukai,~K., et al.: \aap \, {\bf 437}, 935 (2005)

\bibitem[\protect\citeauthoryear{Eastman et al.}{2010}]{eastman10}
Eastman,~J.,  Siverd,~R., Gaudi,~ B.~S.: \pasp \, {\bf 122}, 935 (2010)

\bibitem[\protect\citeauthoryear{Ferrario et al.}{1993}]{ferrario93}
Ferrario,~L., Wickramasinghe,~ D.~T., King,~A.~R.: \mnras  \, {\bf 260}, 149 (1993)

\bibitem[\protect\citeauthoryear{Fritz and Bruch}{1998}]{fritz98}
Fritz,~T., Bruch,~A.: \aap \, {\bf 332}, 586 (1998)

\bibitem[\protect\citeauthoryear{G\"{a}nsicke et al.}{2005}]{gansicke05}
G\"{a}nsicke,~B.~T., Marsh,~T.~R., Edge,~A., et al.: \mnras \, {\bf 361}, 141 (2005)

\bibitem[\protect\citeauthoryear{Halpern et al.}{2018}]{halpern18}
Halpern,~ J.~P., Thorstensen,~J.~R., Cho,~P., Collver,~G., Motsoaledi,~M., et al.: \aj \, {\bf 155}, id. 247 (2018)

\bibitem[\protect\citeauthoryear{Hellier}{1995}]{hellier95}
Hellier,~C.:  In: Buckley,~D.~A.~H., Warner,~B. (eds.) Cape Workshop on Magnetic Cataclysmic Variables. Publ. Astron. Soc. Pac. Conf. Series, vol. 85, p. 185 (1995)

\bibitem[\protect\citeauthoryear{Hellier}{1999}]{hellier99}
Hellier,~C.: \apj \, {\bf 519}, 324 (1999) 

\bibitem[\protect\citeauthoryear{Hellier}{2001}]{hellier01}
Hellier,~C.: Cataclysmic Variable Stars, Springer, 2001

\bibitem[\protect\citeauthoryear{Hellier et al.}{1991}]{hellier91}
Hellier,~C., Cropper,~M., Mason,~K.~O.: \mnras \, {\bf 248}, 233 (1991) 

\bibitem[\protect\citeauthoryear{Kozhevnikov}{2001}]{kozhevnikov01}
Kozhevnikov,~V.~P.: \aap \, {\bf 366}, 891 (2001)

\bibitem[\protect\citeauthoryear{Kozhevnikov}{2002}]{kozhevnikov02}
Kozhevnikov,~V.~P.: In: Battrick,~B., Favata,~F., Roxburgh,~I.~W., Galadi,~D. (eds.) Proceedings of the First Eddington Workshop on Stellar Structure and Habitable Planet Finding, 11 -- 15 June 2001, C\'{o}rdoba, Spain. ESA SP-485, Noordwijk: ESA Publications Division, p. 299 (2002)

\bibitem[\protect\citeauthoryear{Kozhevnikov}{2010}]{kozhevnikov10}
Kozhevnikov,~V.~P.: Astron. Letters \, {\bf 36}, 554 (2010)

\bibitem[\protect\citeauthoryear{Kozhevnikov}{2012}]{kozhevnikov12}
Kozhevnikov,~V.~P.: \mnras \, {\bf 422}, 1518 (2012)

\bibitem[\protect\citeauthoryear{Kozhevnikov}{2014}]{kozhevnikov14}
Kozhevnikov,~V.~P.: \mnras \, {\bf 443}, 2444 (2014)

\bibitem[\protect\citeauthoryear{Kozhevnikov}{2015}]{kozhevnikov15}
Kozhevnikov,~V.~P.: Nev Astron. \, {\bf 41}, 59 (2015)

\bibitem[\protect\citeauthoryear{Kozhevnikov}{2016}]{kozhevnikov16}
Kozhevnikov,~V.~P.: \apss \, {\bf 361}, 273 (2016) doi:10.1007/s10509-016-2859-0

\bibitem[\protect\citeauthoryear{Kozhevnikov}{2017}]{kozhevnikov17}
Kozhevnikov,~V.~P.: \apss \, {\bf 362}, id. 144 (2017) doi:10.1007/s10509-017-3129-5

\bibitem[\protect\citeauthoryear{Kozhevnikov}{2018}]{kozhevnikov18}
Kozhevnikov,~V.~P.: \apss \, {\bf 363}, id. 130 (2018) doi:10.1007/s10509-018-3351-9

\bibitem[\protect\citeauthoryear{Kozhevnikov and Zakharova}{2000}]{kozhevnikoviz}
Kozhevnikov,~V.~P., Zakharova,~P.~E.: In: Garzon,~F., Eiroa,~C., de~Winter,~D.,  
Mahoney,~T.~J. (eds.) Disks, Planetesimals and  Planets. Publ. Astron. Soc. Pac. Conf. 
Series, vol. 219, p. 381 (2000) 

\bibitem[\protect\citeauthoryear{Kuulkers et al.}{2006}]{kuulkers06}
Kuulkers,~E., Norton,~A., Schwope,~A., Warner,~B.: In: Lewin,~W., van~der~Klis,~M. (eds.) Compact stellar X-ray sources. Cambridge Astrophysics Ser., No. 39. Cambridge University Press, Cambridge, p. 421 (2006)

\bibitem[\protect\citeauthoryear{la Dous}{1994}]{ladous94}
la~Dous,~C.: Space Sci. Rev.  {\bf 67}, 1 (1994)

\bibitem[\protect\citeauthoryear{Patterson}{1994}]{patterson94}
Patterson,~J.: \pasp \, {\bf 106}, 209 (1994)

\bibitem[\protect\citeauthoryear{Scaringi}{2014}]{scaringi14}
Scaringi,~S.: \mnras \, {\bf 438}, 1233 (2014)

\bibitem[\protect\citeauthoryear{Schwarzenberg-Czerny}{1991}]{schwarzenberg91}
Schwarzenberg-Czerny,~A.: \mnras \, {\bf 253}, 198 (1991)

\bibitem[\protect\citeauthoryear{Schwarzenberg-Czerny}{1998}]{schwarzenberg98}
Schwarzenberg-Czerny,~A.: Baltic Astron. {\bf 7}, 43 (1998)

\bibitem[\protect\citeauthoryear{Warner}{1986}]{warner86}
Warner,~B.: \mnras \, {\bf 219}, 347 (1986)

\bibitem[\protect\citeauthoryear{Warner}{1995}]{warner95}
Warner,~B.: Cataclysmic Variable Stars. Cambridge Astrophys. Ser., vol. 28. Cambridge University Press, Cambridge (1995)

\bibitem[\protect\citeauthoryear{Warner}{1996}]{warner96}
Warner,~B.: \apss\, {\bf 241}, 263 (1996)








\end{thebibliography}
\end{document}